\documentclass[pdflatex,sn-mathphys-num,iicol]{sn-jnl}%

\usepackage{graphicx}%
\usepackage{multirow}%
\usepackage{amsmath,amssymb,amsfonts}%
\usepackage{amsthm}%
\usepackage{mathrsfs}%
\usepackage[title]{appendix}%
\usepackage{xcolor}%
\usepackage{textcomp}%
\usepackage{manyfoot}%
\usepackage{booktabs}%
\usepackage{algorithm}%
\usepackage{algorithmicx}%
\usepackage{algpseudocode}%
\usepackage{listings}%
\usepackage{amsmath,amssymb}
\usepackage{siunitx}
\usepackage{chemformula}

\theoremstyle{thmstyleone}%

\theoremstyle{thmstyletwo}%

\theoremstyle{thmstylethree}%

\unnumbered

\usepackage{blindtext}
\begin{document}


\title[Article Title]{Photonic-integrated quantum sensor array for microscale magnetic localisation}

\author*[1]{\fnm{Hao-Cheng} \sur{Weng}}\email{haocheng.weng@bristol.ac.uk}

\author[1]{\fnm{John G.} \sur{Rarity}}

\author[1]{\fnm{Krishna C.} \sur{Balram}}

\author*[2,1]{\fnm{Joe A.} \sur{Smith}}\email{joe.a.smith@sheffield.ac.uk}

\affil[1]{\orgdiv{Quantum Engineering Technology Labs, H. H. Wills Physics Laboratory and Department of Electrical and Electronic Engineering}, \orgname{University of Bristol}, \postcode{Bristol BS8 1TL}, \country{United Kingdom}}
\affil[2]{\orgdiv{School of Electrical and Electronic Engineering}, \orgname{University of Sheffield}, \postcode{Sheffield S1 3JD}, \country{United Kingdom}}

\abstract{Nitrogen-vacancy centres (NVs) are promising solid-state nanoscale quantum sensors for applications ranging from material science to biotechnology. Using multiple sensors simultaneously offers advantages for probing spatiotemporal correlations of fluctuating fields or the dynamics of point defects. In this work, by integrating NVs with foundry silicon-nitride photonic integrated circuits, we realise the scalable operation of eight localised NV sensors in an array, with simultaneous, distinct readout of the individual sensors. Using the eight NV sensors and machine-learning methods for multi-point magnetic field reconstruction, we demonstrate microscale magnetic localisation of a 30 \textmu m-sized needle tip. Experimentally, the needle tip can be localised with an error below its dimension and tracked dynamically with high fidelity. We further simulate the feasibility of our platform for monitoring the position and orientation of magnetic microrobots designed for biological and clinical purposes. Without the complexity of bulk optics, our photonic-integrated multi-sensor platform presents a step towards real-life biomedical applications under out-of-the-lab conditions.}


\maketitle


The spin systems of nitrogen-vacancy centres in diamond (NVs) are nanoscale quantum sensors that are optically controllable, biologically compatible, and operable at room temperature. Quantum sensing with NVs has shown impact in material science through condensed matter studies \cite{rovny2024nanoscale, casola2018probing} and biomedical applications through bioimaging and sub-cellular sensing \cite{aslam2023quantum, balasubramanian2014nitrogen}. As opposed to using a single NV sensor probe, multi-NV protocols based on spatially separated single-NV or NV-ensemble sensors have shown promising advantages in resolving target dynamics in both space and time. Some recent demonstrations include the direct imaging of magnetotactic bacteria \cite{le2013optical} and nuclear magnetic resonance
(NMR) signals \cite{briegel2025optical}, the localisation and monitoring of point defect dynamics in materials \cite{ji2024correlated, delord2024correlated}, and the extraction of co-varying electromagnetic fluctuations and spatiotemporal correlations \cite{rovny2022nanoscale, huxter2024multiplexed}.

To realise multi-NV quantum sensing efficiently requires parallelism in experimental hardware for distinct control and readout of each sensor. To scale beyond dual-path confocal microscopy \cite{rovny2022nanoscale}, recent efforts improve conventional widefield spectroscopy by incorporating spatial light modulators (SLMs) \cite{cheng2025massively, cambria2025scalable}. However, widefield ODMR with EMCCD cameras is constrained by frame rates $\geq 10$ ms, which are much slower than the NV spin readout time (300 ns). To improve signal-to-noise, spin-to-charge conversion is often used, providing higher contrast at the cost of additional complexity and longer readout time ($\approx 8$ ms). One can also resolve individual NV sensors by utilising additional degrees of freedom, such as difference in optical resonance \cite{ji2024correlated, le2025wideband} or orientations \cite{huxter2024multiplexed, yoon2025quantum}. However, this poses stringent conditions, such as low-temperature operations or strongly-biased fields, that limit scalability.  

Compared to the bulk-optical methods for multi-NV addressing, photonic-integrated NV platforms uniquely enable scalable, robust, and low-crosstalk operation. Single-mode waveguides and fibre arrays provide per-sensor optical I/O and routing \cite{weng2023heterogeneous}, suppressing stray fluorescence and improving channel isolation; edge-coupled photonic integrated circuits (PICs) support compact packaging with mechanically stable alignment which we find remains stable for weeks (see Methods). Crucially, PIC multi-channel read out through single-photon detectors would avoid camera frame-rate bottlenecks in widefield spectroscopy and confocal scan overheads, allowing sub-nanosecond resolution, and is inherently compatible with time-multiplexed readout \cite{cambria2025scalable} towards high-density sensor grids. 

In this work, we showcase multi-NV quantum sensing with foundry silicon nitride PICs. Parallel operation of eight individually-addressable NV ensemble sensors in an array is enabled by dedicated excitation fibre channels and collection waveguides. By performing Continuous-Wave Optically-Detected-Magnetic-Resonance (CW ODMR) measurements, the eight sensors record spatially-resolved magnetic fields simultaneously, allowing multi-point field reconstruction and direct application in magnetic localisation. With the help of machine learning methods, we can estimate the static position of a 30-\textmu m magnetised needle tip (with an error below the tip dimension) and dynamically track the tip movement to reconstruct its trajectory at minute-level frame rates. We further show in simulation that our NV ensemble sensor array can be used to track magnetic microrobots \cite{jiang2024magnetic} and, with single NV sensors, both the position and the orientation can be estimated precisely, with averaged error of 10 \textmu m and $5^\circ$. Our NV-based magnetic localisation technique (without the need for direct optical access to the target) has an advantage over optical microscopic monitoring in complex environments such as low-light in vivo studies \cite{wrede2022real} or microfluidic channel \cite{jeon2021magnetically} experiments where visibility is limited. The monolithic chip-based platform also unlocks the application of multi-NV schemes for robust biosensing in real-life scenarios.

\begin{figure*}[htp!]
\centering
\includegraphics[width=1\textwidth]{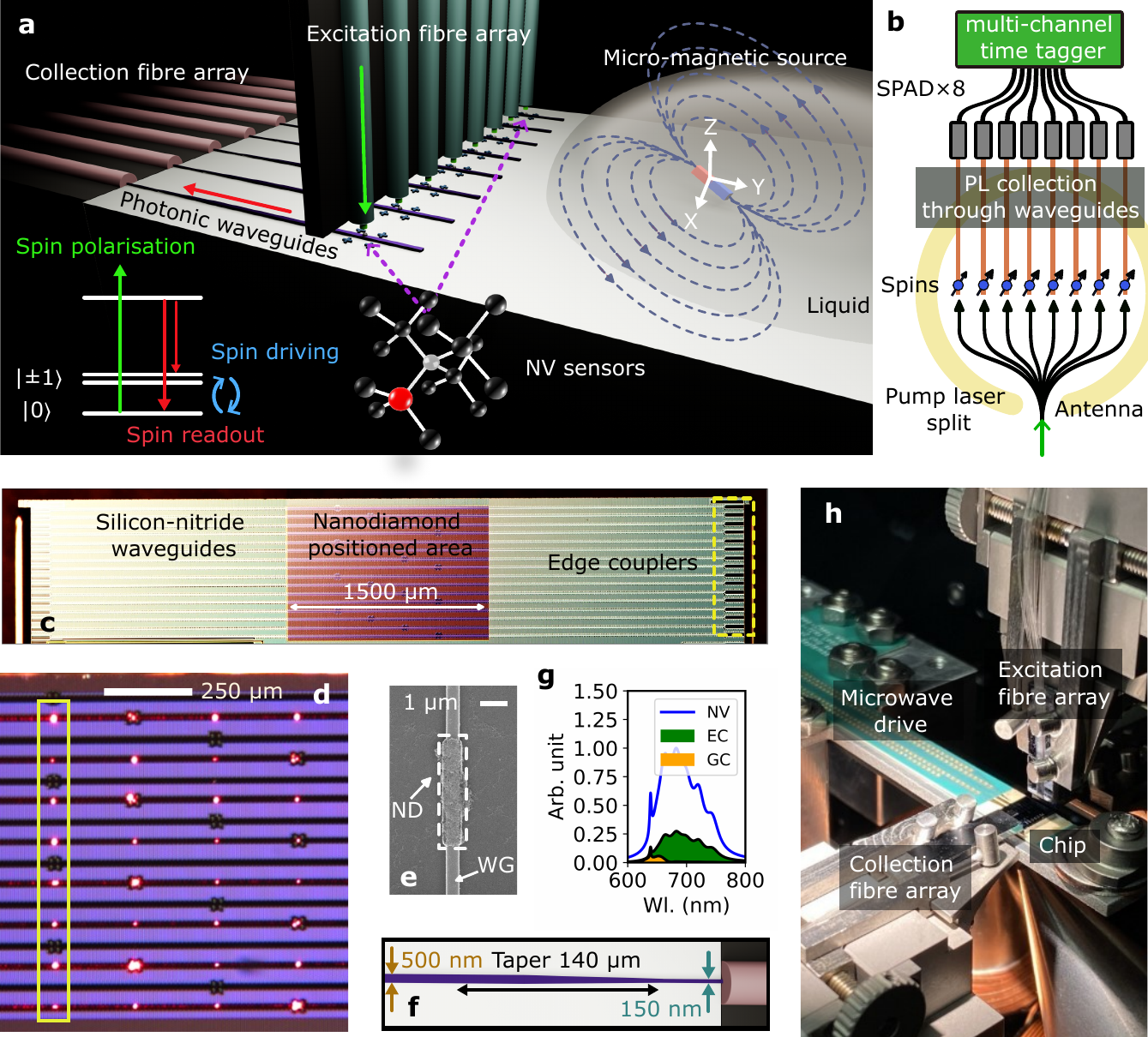}
\caption{\textbf{Multi-NV quantum sensing platform using photonic integrated circuits}. \textbf{a} A schematic representation of multi-NV quantum sensing using photonic integrated circuits. NVs on the photonic waveguides are pumped through the top excitation fibre array and distinctly read out with PL coupled off-chip through the collection fibre array. The eight sensors are used collectively to magnetically localise the position of a microscale magnetic source. Enabled by the high-index-contrast photonic channel, magnetic localisation remains feasible when direct top-down visibility is limited (for example, obscured by a liquid). \textbf{b} Schematic diagram of the experimental setup. Eight NV-spin sensors are pumped by a 515 nm laser split into eight channels. The distinct PL signal is collected by parallel single-mode waveguides and sent to SPAD for detection. The eight SPAD signals are sent to a multi-channel time tagger for spin-state measurements. The NV spins are driven by a global microwave antenna (yellow ring in the diagram). \textbf{c} The silicon-nitride photonic chip. Experimentally, we use only eight out of the sixteen waveguides in parallel. The middle read area is unclad for nanodiamond positioning. The edge couplers (enclosed by the yellow dashed-line box) are fabricated with a deep etch to ensure surface smoothness. \textbf{d} Image of a nanodiamond array. Here, the nanodiamonds are visualised by scattered laser light when a red laser is coupled into the waveguides. Scattered spots look extended due to scattering of light around the alignment markers. \textbf{e} A scanning-electron-microscopic image of a lithographically-positioned nanodiamond site, showing a cluster of nanodiamonds in the rectangular window. \textbf{f} The schematic diagram of waveguide tapering to the edge couplers. The 150 nm end size is limited by the foundry minimum feature size. \textbf{g} Comparison of the coupling efficiency over the NV spectrum (blue) for edge couplers (green) and grating couplers (yellow). Wavelength abbreviated Wl. in the plot. \textbf{h} Experimental setup, showing the excitation and collection fibre array, the chip, and microwave delivery through a PCB antenna.
}\label{fig1}
\end{figure*}

\subsection{Scalable NV quantum sensing with photonic integrated circuits}\label{sec_idea}

\begin{figure*}[ht!]
\centering
\includegraphics[width=\textwidth]{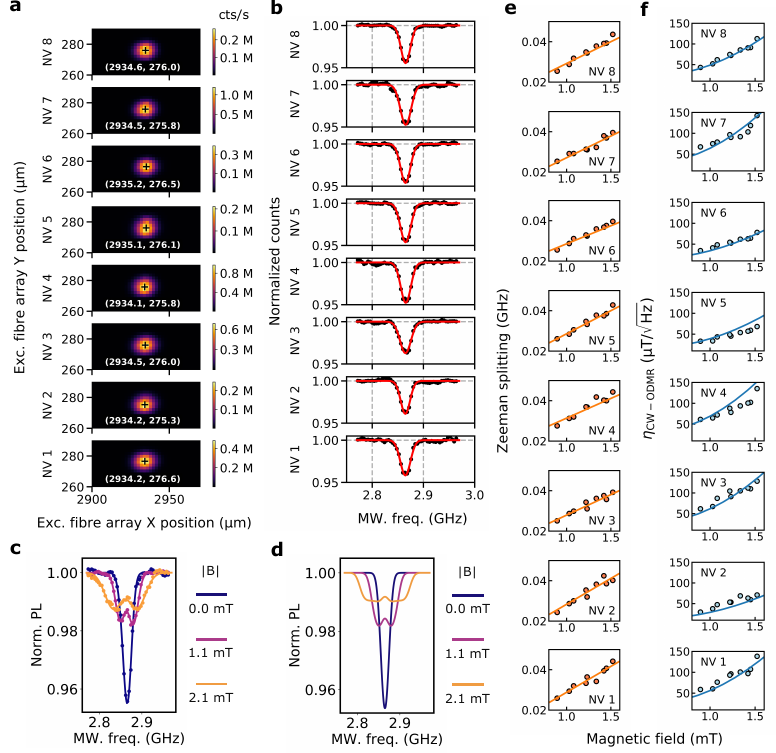}
\caption{\textbf{Parallel sensor operation and characterisation}. \textbf{a} The PL signal of the eight NV sensors collected from each waveguide when the excitation (exc.) fibre array is scanned. Here, the X direction is parallel to the waveguides. Fitted centres of the Gaussian spots are labelled in \textmu m. The colour bar unit is counts per second. \textbf{b} The CW ODMR measurements of the eight NV sensors (NV 1-8) in parallel under zero field. Microwave frequency is abbreviated as MW. freq. \textbf{c} Experimental Zeeman-splitting showing the NV ensemble response. This corresponds to simulations in \textbf{d}. By fitting the resonance with Gaussian dip(s), the FWHM width and contrast can be extracted. \textbf{e} Characterisation of the Zeeman splitting for the eight sensors. Linear splitting is found with respect to the external magnetic field strength. \textbf{f} Characterisation of the CW-ODMR sensitivity as a function of magnetic field strength. The insensitivity (increased $\eta_\textrm{CW-ODMR}$) under higher field is due to resonance broadening and reduced contrast. }\label{fig2}
\end{figure*}

To operate multiple NV quantum sensors at the same time, parallel photonic channels for each sensor are required, for efficient pumping (typically 532 nm or 515 nm laser) to polarise the spins and broadband (from 600 nm to 800 nm) photoluminescence (PL) collection to distinctly read out the spins. We propose a scalable NV sensor control scheme with photonic integrated circuits and parallel fibre arrays, as shown in Fig. \ref{fig1}a. NV centres in nanodiamonds allow deterministic positioning of the sensors on the silicon nitride photonic waveguides. The NV centres are pumped from the top by 515 nm laser through an 8-channel single-mode-fibre array. The PL signal is evanescently coupled to the single TE-mode waveguides \cite{weng2023heterogeneous} and sent off-chip to a second 8-channel collection fibre array through edge couplers. The setup schematic is shown in Fig. \ref{fig1}b. Note that by coupling the NV dipole emission to single mode waveguides, this helps reject fluorescence from the excitation fibre \cite{weng2023heterogeneous} and presents superior NV-emission collection efficiency compared to multimode-fibre end-scope architectures \cite{wojciechowski2019optical}. 

We realise this architecture with a commercial silicon-nitride photonic chip (see Methods), consisting of sixteen parallel waveguides as shown in Fig. \ref{fig1}c. The central parts of the waveguides have the top silica cladding removed and sites labelled with alignment markers, allowing NV centres in nanodiamonds to be heterogeneously combined \cite{weng2023heterogeneous}. In Fig. \ref{fig1}d, we show the deterministic positioning of 100-nm-sized fluorescent nanodiamonds (FND) in an array through a lithographic deposition method \cite{weng2023heterogeneous, tiwari2025single, weng2025crosstalkmitigatedmicroelectroniccontrolopticallyactive} (see also Methods). The FNDs, each hosting more than 1000 NVs, are selected to guarantee near-unity NV yield per nanodiamond site and provide a strong optical signal. Note that each spot contains a cluster of FNDs in a 4 \textmu m by 0.5 \textmu m area (Fig. \ref{fig1}e), chosen to match the mode field diameter of the excitation fibre. The yellow box in Fig. \ref{fig1}d shows the eight NV sensors used for this work. For edge coupling, the waveguides are tapered from 500 nm to 150 nm in 140 \textmu m at the end for mode matching (see Fig. \ref{fig1}f). With broadband operability, we measure a nine-fold enhancement in coupling efficiency over the NV spectrum compared to previous works with grating couplers \cite{weng2023heterogeneous} (see Fig. \ref{fig1}g). The in-plane coupling also presents an advantage over grating couplers for packaging stability.

In Fig. \ref{fig1}h, we show the experimental setup  (see also Methods), where a microwave signal is delivered through a Printed Circuit Board (PCB) antenna (underneath the chip) to drive the NV spins in the array collectively. The PL signal from the eight collection fibre channels is sent to optical filtering and single photon avalanche diodes (SPADs) for detection. Note that by replacing the 8-channel excitation fibre array with an eight-by-eight matrix fibre array, the full two-dimensional NV array could be used for sensing at the same time (i.e., for the same interrogation time) with spin readout interleaved in time \cite{cambria2025scalable}. Also, the setup generally applies to NV sensors based on single NVs or NV ensembles (see Supplementary Section C for more details).

\subsection{Parallel operation of eight NV magnetic sensors}\label{sec_fab}
Parallel operation of the eight NV sensors is achieved only when the excitation fibre is aligned to the NV array and all eight collection fibres are well coupled to the waveguide channels. Through scanning of the excitation fibre array while recording the PL signal from each waveguide, we show in Fig. \ref{fig2}a that the NV Gaussian spots are aligned within the fibre mode field diameter. Note that the waveguide-to-chip coupling remains at up to 85\% efficiency with eight channels coupled in parallel. While the spot sizes are limited by the fibre mode field diameter and the distance between the excitation fibre and the chip (which is kept within a few \textmu m), the physical size of the sensor (and thus the ultimate spatial resolution for quantum sensing) is defined directly by the deposition area (which is 4 \textmu m by 500 nm). The brightness of each spot varies due to differences in the number of nanodiamonds on each spot, with a minimum of 200k counts per second. 


\begin{figure*}[htbp!]
\centering
\includegraphics[width=1\linewidth]{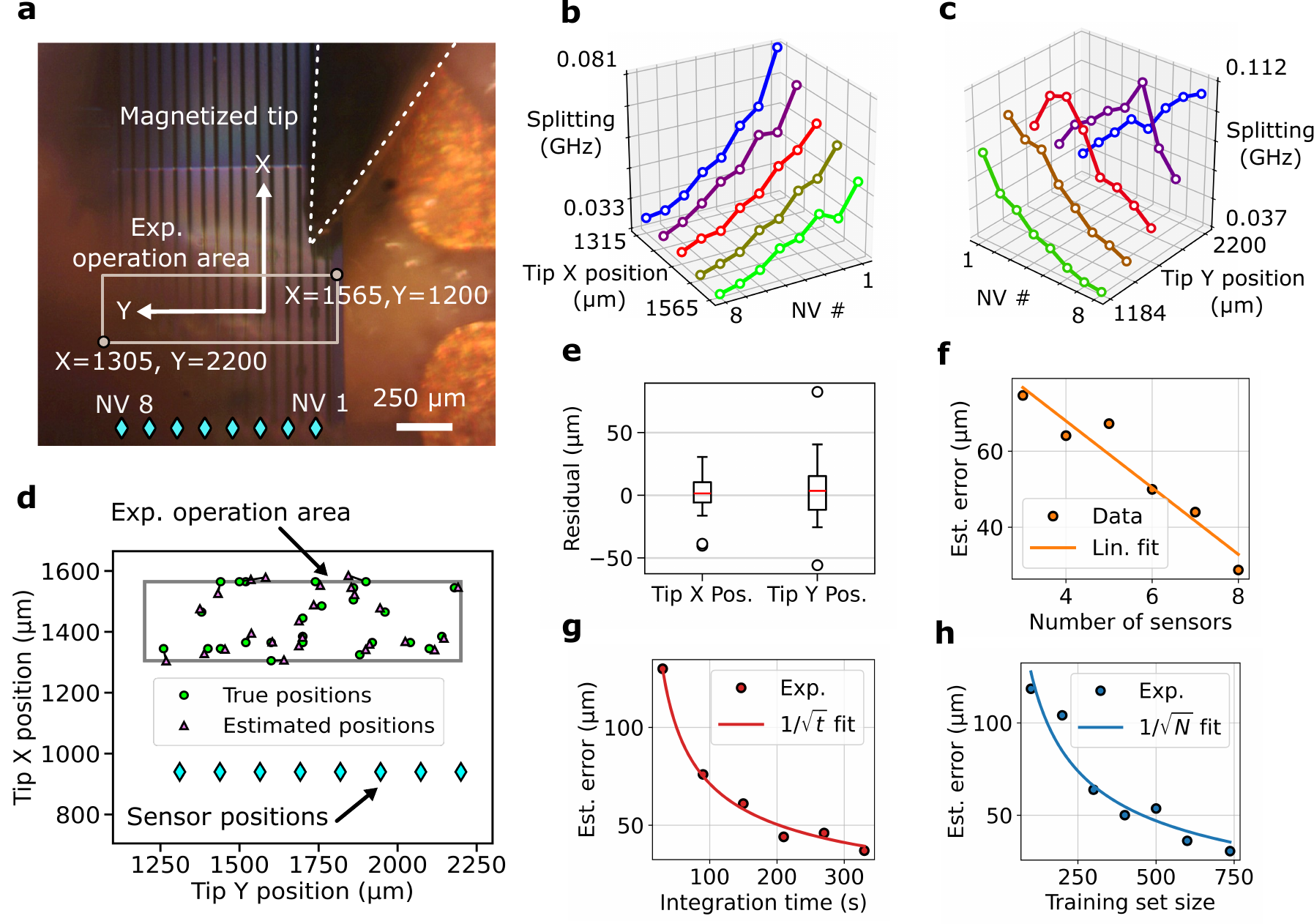}
\caption{\textbf{Experimental magnetic localisation with multi-point field reconstruction.} \textbf{a} Image showing the magnetic needle tip above the chip, the experimental (exp.) operation area, and the sensor relative positions. Note that the same X-Y coordinates (the needle tip motor stage reference frame) in \textmu m are used throughout this figure. \textbf{b} The Zeeman splittings of each NV sensor when the tip moves in the X direction, showing increased gradient. \textbf{c} The Zeeman splittings recorded when the tip moves in the Y direction showing the translation in peak field position. \textbf{d} Experimental magnetic localisation. The sensor array is used to estimate the needle position in X and Y. The experimental operation area is enclosed by the grey box with the sensor positions labelled. \textbf{e} Residual analysis for the magnetic localisation result in \textbf{d}, comparing the needle tip X and Y positions. \textbf{f} The linear decrease of averaged estimation error with increased number of sensors used simultaneously. \textbf{g} The $1/\sqrt{t}$ scaling of the average estimation error with the integration time $t$. All eight sensors are used here. \textbf{h} Relationship between average estimation error and training dataset size $N$, showing $1/\sqrt{N}$ scaling. All eight sensors are in use here with a 360 s integration time.} \label{fig3}
\end{figure*}

We perform CW ODMR measurements of the eight sensors at the same time. With no external field (Fig. \ref{fig2}b), the ODMRs show resonances (fitted to Gaussian dips) with full width at half maximum (FWHM) $\leq25$ MHz and contrasts ($C = (I_{\mathrm{off}} - I_{\mathrm{on}})/I_{\mathrm{off}}$ where $I_{\mathrm{off,on}}$ are PL intensities off/on resonance) $\approx 4\%$ under 5 mW laser power per site and 31 dBm microwave driving power at the source. As each sensor is composed of clusters of randomly oriented nanodiamonds, the sensor effectively becomes a magnetic field magnitude sensor \cite{foy2020wide}. An example set of CW ODMR measurements under an external field is shown in Fig. \ref{fig2}c. With the ensemble effect, our sensors are more sensitive in the low field regime due to decreased ODMR contrast and increased resonance linewidth under a larger field. This matches the simulations in Fig. \ref{fig2}d (see details in Methods) and previously reported results \cite{foy2020wide}. 

We characterise the operation of the magnetic magnitude sensors. For a DC magnetic field in the range 0.8 mT $\leq |B|\leq$ 1.6 mT, we show that the Zeeman splitting is proportional to the field magnitude, Fig. \ref{fig2}e. The solid lines present a linear fit of the data points. The external magnetic field is generated by a strong neodymium magnet 20 centimetres away (pointing along the waveguide direction) to ensure field homogeneity across the eight sensors. The field range measured is only a subset of the full operation range and is selected for experimental convenience. The average slope is $22\pm3$ MHz/mT, which is about half the maximum splitting ($2\times28$ MHz/mT) due to the averaged response of isotropic NV orientations. The magnetic field sensors are characterised by the CW ODMR sensitivity (units: \textmu T/$\sqrt{\textrm{Hz}}$) \cite{barry2020sensitivity}
\begin{equation}
\eta_{\mathrm{cw-ODMR}} = \frac{4}{3 \sqrt{3}} \frac{h}{ \, g_e \ \mu_\textrm{B}} \, \frac{\Delta\nu}{C_{\mathrm{cw}} \  \sqrt{R}},
\end{equation} 
where $h$ is the Planck constant, $g_e$ the electron g-factor, $\mu_\textrm{B}$ is the Bohr magneton, $\Delta\nu$ is the resonance linewidth (FWHM), $C_{\mathrm{cw}}$ is the CW-ODMR contrast, and $R$ is the photon-detection count rate. The $\eta_{\mathrm{cw-ODMR}}$ can also be understood as the minimum DC magnetic field we can detect using CW-ODMR with 1 Hz bandwidth and the $\sqrt{\textrm{Hz}}$ scaling indicates that the sensitivity improves (the minimum detectable field decreases) with the square root of integration time. We characterise the magnetic field sensitivity of the eight sensors in Fig. \ref{fig2}f. Note that in zero field, the sensitivity $\approx25$ \textmu T/$\sqrt{\textrm{Hz}}$ is similar to other NV-ensemble-based sensors \cite{foy2020wide, kim2019cmos}. Under 1 mT magnetic field strength, $\eta_{\mathrm{cw-ODMR}}$ increases to $\geq 50$ \textmu T/$\sqrt{\textrm{Hz}}$ and doubles again to $\geq100$ \textmu T/$\sqrt{\textrm{Hz}}$ under 1.5 mT. The experimental data points are compared with extracted values from fitted ensemble sensor models (solid lines). 

The full operating window of the sensors, however, extends from 0.2 mT to 2.2 mT (see below and Supplementary Section A). For $|B|<$ 0.2 mT, the Zeeman splitting is insensitive to external fields due to strain in the nanodiamond, while the ODMR contrast decreases significantly to $\leq1\%$ for $|B|>$ 0.2 mT. Compared with single-NV CW-ODMR sensitivity around 1 \textmu T/$\sqrt{\textrm{Hz}}$ \cite{dreau2011avoiding} and NV-ensemble sensitivity around 10 nT/$\sqrt{\textrm{Hz}}$ \cite{zhang2020high}, the 10-100 \textmu T/$\sqrt{\textrm{Hz}}$ sensitivity reported here has room for optimisation via improved diamond quality, higher pump power, and advanced magnetometry protocols (see Conclusions). As the sensitivity of the sensor scales with $1/\sqrt{R}$, and we operate with optical power far below saturation, one can significantly increase the sensitivity (decrease $\eta_{\textrm{CW-ODMR}}$) by pumping the sensors much harder and detecting the signal with low-noise photodiodes as opposed to single-photon detectors. 

\begin{figure*}[htbp!]
\centering
\includegraphics[width=1\linewidth]{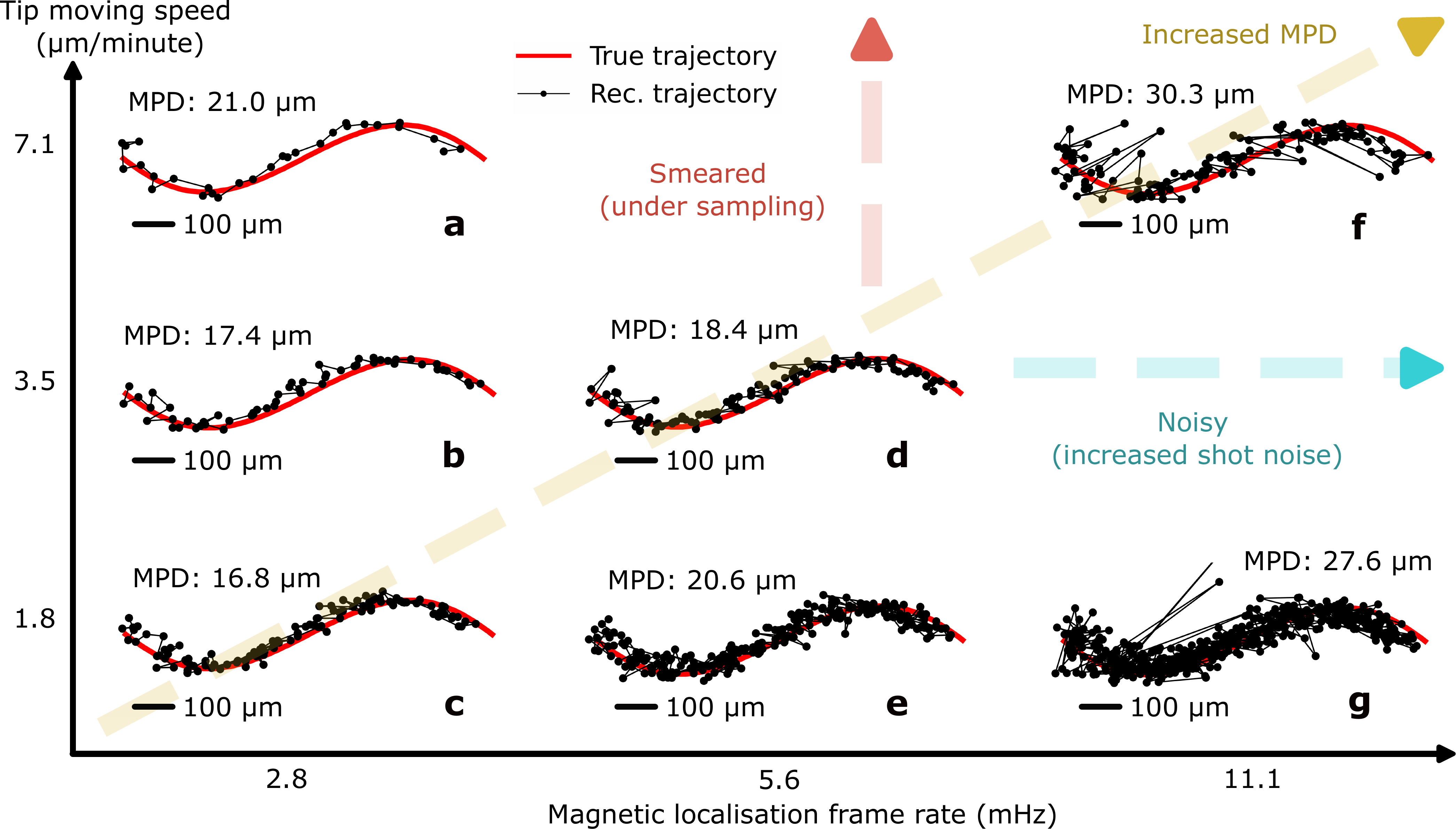}
\caption{\textbf{Experimental dynamical tracking of moving magnetic object}. The needle tip, when moving at different speeds, is tracked with different frame rates for comparison. Note that the same machine-learning model is used for magnetic localisation and the trajectory falls within the operation area in Fig. \ref{fig3}a. The deviation of the reconstructed (rec.) trajectory from the true trajectory is quantified by the Mean Perpendicular Distance (MPD), which increases with higher frame rate due to shot noise and higher moving speed due to undersampling.
} \label{fig4}
\end{figure*}

\subsection{Static magnetic localisation through multi-point field reconstruction}\label{sec_result}

By operating the eight spatially resolved magnetic magnitude sensors at the same time, we first show imaging of a gradient field. The gradient field (at few \textmu T/\textmu m level) is generated by a magnetised needle tip placed close to the sensors (Fig. \ref{fig3}a). The needle tip is 30 \textmu m in diameter, kept at a distance of 250 \textmu m above the chip, and tilted at a 45$^\circ$ angle (see Supplementary Section B). When the tip is directed towards NV sensor 1 (NV 1), moving the tip closer (in the negative X direction of Fig. \ref{fig3}a), increases field gradient across the sensors (Fig. \ref{fig3}b). Conversely, when the tip moves across the waveguides (in the positive Y direction of Fig. \ref{fig3}a) while maintaining its distance from the sensors, one can see the change in peak field location, moving from NV sensor 1 (NV 1) to NV sensor 8 (NV 8) in Fig. \ref{fig3}c. The ability to image a gradient field in real time implies that the eight sensors are recording distinct and localised magnetic fields in parallel, and can be used for multi-point field reconstruction.

Through multi-point field reconstruction, the relations between sensor readouts can be used to estimate the needle tip position from a magnetic inverse calculation, i.e., to implement magnetic localisation. Localising a DC/AC magnetic source is a core technique in many bioimaging \cite{flynn2005biomagnetic} and engineering \cite{pasku2017magnetic} methods. We utilise a Convolutional Neural Network (CNN)-based method (see details in Methods) for the magnetic inverse calculation. The machine learning method, also commonly used in biomedical magnetic inverse imaging \cite{jin2017deep} and geophysical magnetic source localisation \cite{puzyrev2019deep}, is chosen for generalisation compared to simple field inversion, especially when the magnetic field distribution of the source is unknown or hard to fully characterise. Also, the CNN-based model, in contrast to a simple deep neural network, can better capture the relation between spatially-structured sensor readouts for comparison. The model is trained with 737 labelled experimental data points, sampled across an area of 260 \textmu m by 1000 \textmu m (the gray box in Fig. \ref{fig3}a), to learn the X-Y position of the needle tip by the Zeeman splittings of the eight sensors along with the standard deviation of fitted resonance frequencies in the ODMR measurements. The centre of the area is chosen about 500 \textmu m from the sensors such that the field gradient is strong enough to show distinct trends (as in Fig. \ref{fig3}b and c) but the field strength stays within the low field regime to operate with reasonable sensitivities.

In Fig. \ref{fig3}d, we test the trained CNN model with data points unused in the training process, comparing the estimated positions (pink triangular data points) with the true positions of the needle tip (cyan circular data points). The data points are paired up with connection lines to show the estimation error. The operation area and the sensor positions are also labelled in the same figure for reference. We observe an average of 23 \textmu m estimation error, which falls within the tip size of 30 \textmu m (see Supplementary Section B). The position estimation shows a slightly larger error in Y positions compared to the X positions since the magnetic gradient is smaller in the Y direction (see Supplementary Section B). Fig. \ref{fig3}e shows the X and Y residual with an interquartile range of 16 \textmu m and 27 \textmu m respectively. The outliers are cases where a misalignment in the setup caused drops in the PL signal. Note that for each data point here, we integrate the ODMR measurements for 360 seconds.

Using more sensors for magnetic localisation allows us to collect more data simultaneously, which improves the reconstruction of the magnetic field. This advantage is captured by Fig. \ref{fig3}f, showing the estimation error reduces linearly with the number of sensors used. By integrating the ODMR measurements for a shorter period of time, the localisation estimations can be made faster, however, at a higher noise level. We show the scaling between integration time $t$ and estimation error in Fig. \ref{fig3}g, which matches a $1/\sqrt{t}$ fitting. To see how much training data is needed for machine-learning based magnetic localisation, we analyse how the averaged estimation error scales with the dataset size $N$ in Fig. \ref{fig3}h. The estimation error follows $1/\sqrt{N}$ scaling. This is expected since, with fixed operation area $A$, $\sqrt{A/N}$ presents the resolution of sampled data points (with length unit).

The position estimation error is determined by both the sensor sensitivity and the local magnetic field gradient. To understand the relation, we model the spatially-averaged error $\overline{R}_{x,y}$, by averaging the position-dependent error ${R}_{x,y}$ over the sample area where
\begin{equation}
R_{x,y} \approx \frac{S_{x,y}}{\sqrt{t} \ m_{x,y}}.
\label{eq_res}
\end{equation}
Here, $S_{x,y}$ is the sensor sensitivity at position $(x,y)$, $t$ is the integration time for the ODMR measurements, and $m_{x,y}$ is the field gradient generated by the magnetic source. From the sensor sensitivities in Fig. \ref{fig2}f, the integration time (360 s), and the field gradient generated by the tip (see Supplementary Section B) in the Fig. \ref{fig3}d operation area, we estimate $\overline{R}_{x,y}\approx22.5$ \textmu m, with an averaged error in the x direction of 10.6 \textmu m and 17.9 \textmu m in the y direction. This is in good agreement with the experimentally observed 23 \textmu m averaged error in Fig. \ref{fig3}d and the larger Y position error in Fig. \ref{fig3}e.

\begin{figure*}[htbp!]
\centering
\includegraphics[width=0.8\linewidth]{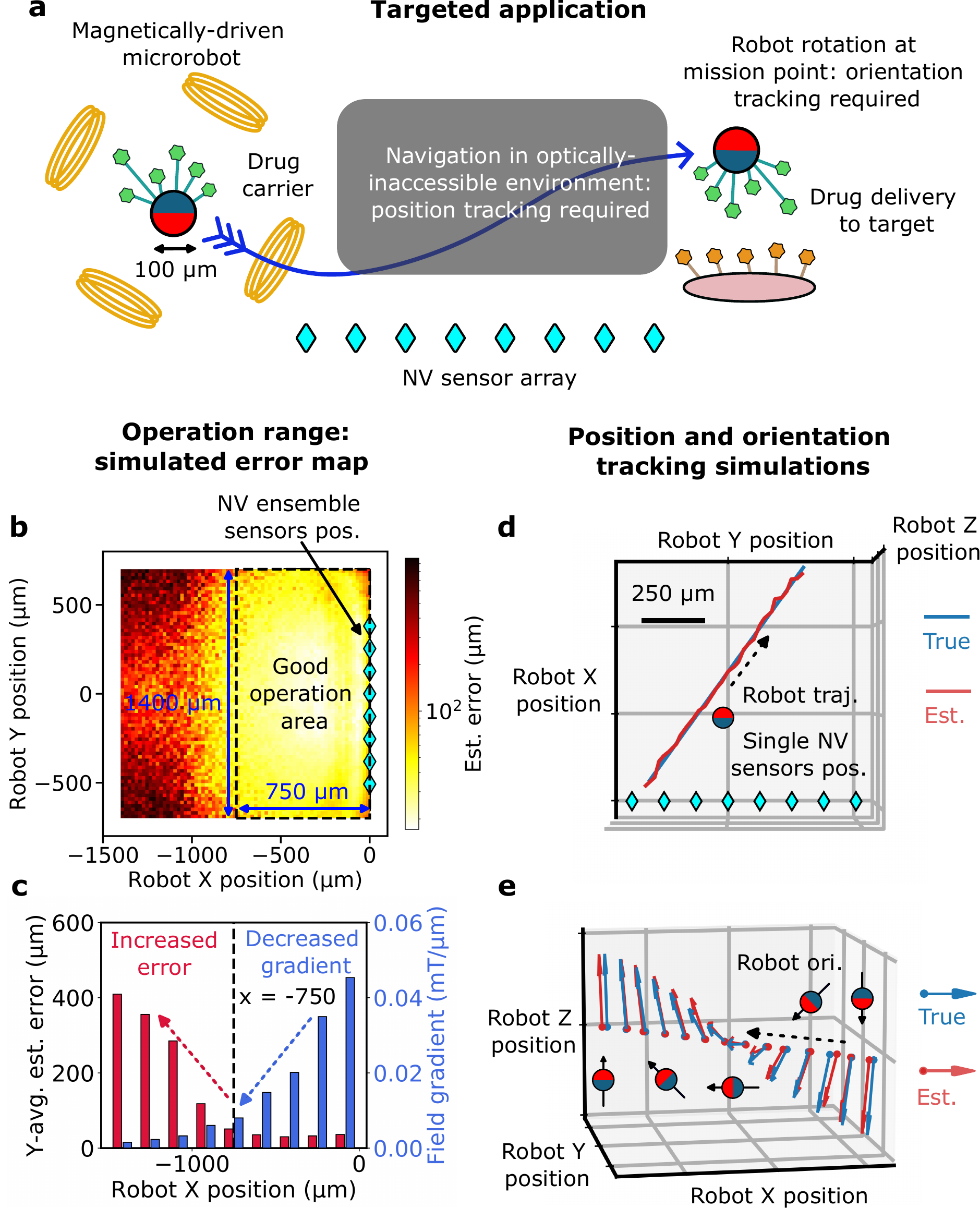}
\caption{\textbf{Application to position and orientation tracking of magnetically-driven microrobot}.  
\textbf{a} Targeted application scenario. The cartoon diagram shows that a 100-\textmu m sized magnetically-driven microrobot, on a mission of pin-point drug delivery, requires the ability to navigate through optically-inaccessible environments with precise control. With our magnetic localisation technique, the microrobot can be efficiently localised and tracked when performing biomedical tasks using an array of NV ensemble or single-NV sensors. Note that the magnetic field used to drive the microrobot is homogeneous across the sensors whose contribution can be easily eliminated in the magnetic localisation calculations. \textbf{b} Error map of simulated microrobot magnetic localisation, with eight ensemble NV sensors. The robot-position-dependent estimation error is plotted with the relative positions of the NV ensemble sensors labelled. The good operation area (with estimation error $\leq50$ \textmu m) is shown by the dashed-line box. \textbf{c} The y-averaged estimation error and the averaged magnetic field gradient at the sensor locations for different robot x positions. When the robot moves away from the sensors, the estimation error increases as a result of decreased field gradient. See also Supplementary Section D for the microrobot magnetic field profile. The dashed vertical line at x=-750 \textmu m shows where the good operation range is defined (when estimation error stays $\leq 50$ \textmu m). \textbf{d} Position tracking of the microrobot along a trajectory (abbreviated as traj.) with eight single NV sensors. The single-NV sensor positions are labelled. We consider the single NV sensors only 150 \textmu m below the microrobot since their superior sensitivity at higher field allows the robot to be tracked with a smaller z standoff. \textbf{e} Monitoring of microrobot rotations. The change of orientations (abbreviated as ori.) alongside the trajectory (as visualised by the robot diagram) can be precisely tracked with the single NV sensors. Arrows in the plot show the true and estimated orientations for every two points along the trajectory.
} \label{fig5}
\end{figure*}

\subsection{Dynamics of a moving magnetic source: continuous-time tracking}\label{sec_result}

Beyond localising stationary magnetic objects, we can use the eight sensors to track a moving target in continuous time. To demonstrate this, we program the motor stage such that the needle tip follows an S-shaped trajectory in 100 steps, with a speed of 7.1 \textmu m/minute, 3.5 \textmu m/minute, 1.8 \textmu m/minute (defined by the 10.6 \textmu m step size and the 90 s, 180 s, and 360 s dwell time at each position). The ODMR measurements of the eight sensors are taken continuously to record the dynamics of the needle tip. The trajectory is then reconstructed with frame rates of 2.8 mHz, 5.6 mHz, and 11.1 mHz (defined by one over the 360 s, 180 s, and 90 s integration time). The results are shown in Fig. \ref{fig4}, with the red curve showing the true trajectory and the connected data points giving the reconstructed trajectory. Comparing Fig. \ref{fig4}c, Fig. \ref{fig4}e, and Fig. \ref{fig4}g, when the needle tip moves at a certain speed, a noisier trajectory is observed at higher frame rates (as expected from Fig. \ref{fig3}g due to shot noise). On the other hand, when the needle tip speed increases and the frame rate is fixed, we observe a smeared trajectory due to undersampling. For example, a larger deviation from the true trajectory can be seen in Fig. \ref{fig4}a, especially where the curvature is large, compared to Fig. \ref{fig4}b and Fig. \ref{fig4}c. To quantify the fidelity of the trajectory reconstruction, we calculate the error in terms of the Mean Perpendicular Distance (MPD), i.e., the averaged shortest distance from each estimated data point to the true trajectory. Note that the stationary averaged estimation error is not used as the metric here because the movement is typically not synchronised with the ODMR measurement integration time in tracking unknown dynamics (i.e. the magnetic localisation can be performed while the tracked object moves, depending on the frame rate chosen). We can see an increased MPD for both more severe undersampling and increased shot noise. MPD increases from the lower left to the upper right of the figure as illustrated. 

\subsection{Magnetic microrobot tracking- simulations and the use of single-NV sensors}\label{sec_result}

Our platform enables (i) a multi-NV sensing architecture without the use of complex bulk optics, and (ii) a microscale localisation technique without need for direct optical access to the target, presenting a practical and deployable solution for real-life quantum sensing applications. Here, we show in simulation one such application: dynamical tracking of in-plane magnetic microrobot dynamics, for many clinical and biomedical purposes \cite{wu2025femtosecond, jiang2024magnetic, han2024janus, wrede2022real}. The concept is shown in Fig. \ref{fig5}a, where our magnetic localisation technique can help navigate a magnetically-driven microrobot in an optically-inaccessible environment for a drug-delivery mission (similar to that in \cite{han2024janus}). 

For microrobot tracking applications, it is crucial to characterise the valid spatial range of operation. We show this by simulation (see details in Supplementary Section D), considering a magnetic hydrogel microrobot made of \ch{Fe3O4} particles \cite{jiang2024magnetic} with a diameter of 100 \textmu m, whose magnetic profile is simulated by COMSOL multiphysics. Note that the simulation considers the NV ensemble quantum sensor sensitivity in Fig. \ref{fig2}, the ODMR integration time of 360 seconds, and photonic layout in Fig. \ref{fig1}. In Fig. \ref{fig5}b, we show the position-dependent error map of the microrobot localisation, calculating the estimation error for positions sampled over a 1400 \textmu m by 1400 \textmu m area. Within a good operation area of 750 \textmu m by 1400 \textmu m (shown by the dashed-line box), we observe $\leq50$ \textmu m estimation error. The estimation error is captured by Eq. \ref{eq_res}-- the magnetic localisation fails when the microrobot moves too far away from the sensors and thus the magnetic field gradient is significantly reduced. This argument is supported by the statistics shown in Fig. \ref{fig5}c, plotting the y-averaged estimation error in different robot X positions. The anti-correlation between the averaged field gradient at the NV sensors and the estimation error is observed. When the gradient falls below 0.008 mT/\textmu m, the localisation error exceeds 50 \textmu m (see Fig. \ref{fig5}c), defining the boundary of the good operation area at x=-750 \textmu m. This gradient threshold sets the maximum tracking distance and depends on both sensor sensitivity and integration time. Note that the decreased $m_{x,y}$ dominates over decreased $S_{x,y}$ here (due to resonance broadening), as the NV ensemble sensors are kept reasonably far away in z with a standoff of 250 \textmu m. Also, this position dependence is not obvious in Fig. \ref{fig3}d as a smaller operational area (span in the Fig. \ref{fig3}a X direction) is explored due to physical constraints in the experimental setup.

We further consider the use of single NVs as quantum sensors integrated with photonic waveguides for magnetic localisation. Single NVs have been integrated in similar platforms \cite{weng2023heterogeneous} and we also show that CW ODMR and coherent spin operation (Rabi oscillations) is possible experimentally (see Supplementary Section C). A single NV sensor, with much weaker optical signal but narrower resonance linewidth, is measured with a 25 \textmu T/$\sqrt{\textrm{Hz}}$ CW ODMR sensitivity under a weak unaligned external field. The single NV sensitivity can degrade under unaligned fields due to spin mixing and quenching \cite{rondin2014magnetometry} but less significantly in the low field regime ($<5$ mT) compared to NV ensembles (as shown in Fig. \ref{fig3}f). Also, positioned single NVs in nanodiamonds naturally have different orientations that can be used collectively to provide vectorial information of the magnetic field. We show in simulation (Fig. \ref{fig5}d) that the position of the same 100-\textmu m microrobot can be tracked by eight single NV sensors, with an average estimation error of $\leq10$ \textmu m (see details in Supplementary Section D). In this simulation, the single NV sensitivity (along with random orientations), the same photonic scheme, and 360 seconds of ODMR integration time are considered. Furthermore, the controlled rotation of the microrobot along the trajectory (Fig. \ref{fig5}e) can be simultaneously monitored, with $\approx 5^\circ$ average estimation error.

\subsection{Conclusions}\label{sec12}

In this work, we introduce multi-NV quantum sensing using foundry silicon-nitride integrated photonic circuits. The uniformly positioned NV sensors are controlled by parallel fibres and readout through distinct waveguides. Enabled by multi-point field reconstruction with the eight NV sensors in an array, we experimentally demonstrate magnetic localisation of a 30 \textmu m needle tip with error below its dimension and dynamic tracking of the tip position. We further present, by simulation, the potential of our platform for position and orientation monitoring of magnetic microrobots for biomedical applications.

The frame-rate requirements ($F$) for tracking moving microrobots are determined by both motion speed ($v$) (to avoid undersampling) and the desired estimation accuracy ($\delta_L$, representing estimation error), that is, $F\geq v/\delta_L$. Considering the 360-second CW-ODMR integration time ($F=1/360$ s=2.8 mHz tracking frame rate) and the $\leq 50$ \textmu m precision ($\delta_L=50$ \textmu m) as shown in Fig. \ref{fig5}b (good operation area), motion speed below 0.14 \textmu m/s is trackable ($50 \times 2.8 \times 10^{-3}$ \textmu m/s). This falls in the range of typical microrobot speeds at 0.1–10 \textmu m/s \cite{jiang2024magnetic, wrede2022real}, showing application feasibility. The trackable angular speed of motion ($\Omega$) can also be estimated by $\Omega \leq F \times \delta_\theta$, where $F$ is the 2.8 mHz frame rate and $\delta_\theta$ is the $5^\circ$ angular precision shown in Fig. \ref{fig5}e. We estimate a $2.4 \times 10^{-4}$ rad/s ($\Omega \leq F \times \delta_\theta = 2.8 \times 10^{-3} \  \textrm{Hz} \times 5^\circ \times \frac{\pi}{180} = 2.4 \times 10^{-4}$ rad/s) angular speed upper limit, which is lower than the 30-120 rad/s angular speed in typical operation \cite{fu2015characteristic, alshafeei2014magnetic}. The tracking speed limit (translation and angular) can be significantly improved by adopting NV-ensemble sensors based on single-crystal diamond membranes, heterogeneously combined and positioned \cite{guo2024direct, wan2020large} to the silicon-nitride photonics. Defined orientation in NV ensembles in single-crystal diamonds would enable vector field sensing and thus microrobot orientation tracking. By improving the sensitivity to around 1 nT/$\sqrt{\textrm{Hz}}$ level \cite{zhang2020high, guo2025enhanced}, the same position/angular estimation precision can be achieved with $10^{-10}=(1 \ \textrm{nT}/100$ \textmu $\textrm{T})^2$ shorter integration time (following the relation in Eq. \ref{eq_res}) and thus tracking frame rates beyond MHz level ($2.8 \times 10^{-3} \times 10^{10}=2.8 \times 10^7$ Hz) . For example, an angular speed limit at $2.4 \times 10^6$ rad/s ($2.8 \times 10^7 \ \textrm{Hz} \times  5^\circ \times \frac{\pi}{180} =2.4 \times 10^6$ rad/s) can be achieved. 

Our platform, based on commercial silicon photonics, demonstrates scalability and stability compared to direct-written diamond waveguides \cite{guo2025enhanced} and single mode fibre assemblies \cite{bopp2025diamond} for out-of-the-lab quantum sensing conditions. The deterministically positioned NVs also allow static control with fibre arrays and waveguides, in contrast to stochastic sensors that need to be addressed by fast programmable optics \cite{cheng2025massively, cambria2025scalable}. Our platform can also be adopted for other multi-NV sensing protocols \cite{rovny2022nanoscale, yoon2025quantum} for different use cases. Moreover, with improved photonic architecture, NV ensembles could be pumped via in-plane waveguide channels, thereby freeing access to the space above the sensors. With complete photonic packaging, chip-based quantum sensing could facilitate clinical and biomedical applications where bulk optical imaging is impractical \cite{boto2018moving, monge2017localization}.

\subsection{Methods}\label{sec11}

\subsubsection*{Experimental setup}
Here we include details of the experimental setup. The photonic chip is fabricated by IMEC (BioPIX silicon nitride 150 nm platform). The fibre arrays used for pumping the NV and collection of PL from both sides are made by Oz Optics and Precision Micro-Optic, with eight PM630 polarisation-maintaining single mode fibres in a V-groove array separated by 127 \textmu m. The excitation fibre array is mounted on a scanning stage with a tip, tilt, and rotation stage (TTR001/M Thorlabs) whereas the collection fibre arrays are mounted on six axis stages (Nanomax Thorlabs). The NV centres are pumped by a 40 mW CW Cobolt 06-MLD laser. The NV spins are driven by a microwave source (SMB100AP20 Rohde \& Schwarz), amplified by microwave amplifiers (AM4-2-6-43-43R Microwave Amplifiers Ltd) and delivered by a PCB grounded coplanar waveguide antenna. The PL signal from eight sensors passes through a notch filter (NF03-514E-25 Semrock) and long pass filter (BLP01-568R-25 Semrock). With a pair of fibre bundles (BF74HS01 Thorlabs), the same optical filters can be shared by up to seven individual beam paths. Through the fibre bundle and filter sets, we measure an averaged 65\% efficiency for PL signal in the transmission band and averaged $<0.1\%$ crosstalk between fibre channels. Eight single photon avalanche diodes (SPCM-AQRH-12-FC Excelitas) are used for PL detection. The timetags are recorded by a Logic-16 (UQDevices) where eight out of sixteen channels are used for data processing at the same time. In ODMR measurements, the Pulse Streamer 8/2 (Swabian Instruments) is used to trigger the microwave source and synchronise with the photon counting. The ODMR is set with a dwell time of 1 ms on each frequency to ensure fast data processing. The excitation fibre array is scanned by a motorised three-axis stage, positioned a few microns away from the chip surface to ensure efficient excitation power density. Note that the photonic coupling (fibre to waveguide) and bulk optics efficiency remain stable for weeks with only minimal tuning needed.

\subsubsection*{Nanodiamond positioning}
The nanodiamond positioning method follows Ref. \cite{weng2023heterogeneous} with minor differences. The photonic chip is first covered by a polymethyl methacrylate (PMMA) mask, on which the deposition sites are written with electron beam lithography (Raith Voyager). After development, this creates precisely defined windows on the waveguide for nanodiamond positioning. The fluorescent nanodiamond (brFND-100 FND Biotech) is suspended in deionised water (at 1mg/ml concentration), followed by 2-hour ultrasonication and filtering with 0.45 \textmu m syringe filters. Drops of the nanodiamond suspension are then pipetted on the deposition area (red middle region in Fig. \ref{fig1}c) and dried at room temperature. The dropcasting is repeated five times to ensure proper coverage over the area. The residual nanodiamonds on the PMMA mask are then removed by acetone and gentle ultrasonication for 10 seconds.

\subsubsection*{NV ensemble simulations}
For NV ensembles in nanodiamonds, we simulate CW ODMR measurements based on \cite{foy2020wide}. For a single NV, the electron spin resonance frequencies are $\nu_{\pm1}=D \pm\sqrt{E^2+(\gamma \vec{B} \cdot \hat{a})^2}$, where $E$ captures the effect of strain in nanodiamond, $\gamma$ is the NV gyromagnetic ratio $\approx 28$ GHz/T, $\vec{B}$ is the external magnetic vector field, and $\hat{a}$ is the unit vector along the NV axis. Each resonance in the CW ODMR can be presented by a Gaussian dip $1-G(\nu, \delta \nu, C)$, where $\nu$ is the central frequency, $\delta \nu$ is the associated resonance linewidth, and $C$ is the contrast. Considering an NV ensemble with isotropic distribution of orientations, the collective response can be modelled by $1-n\int_{0}^{\pi} \Big[G( D-\sqrt{E^2+(\gamma \ |B| \cos{\theta})^2},\delta\nu,C)+G(D +\sqrt{E^2+(\gamma \ |B| \cos{\theta})^2}, \delta \nu,C) \Big]\,\sin{\theta} \ d\theta$, where $\theta$ considers the orientation averaging and $n$ is a normalisation factor. Here, we assume no angular and spin dependence for the NV fluorescence, contrast $C$, strain $E$, and linewidth $\delta \nu$, as well as homogeneity of these parameters in the ensemble. These parameters (with an average of $E=5.6$ MHz, $C=3.4\%$, and $\delta \nu= 14.7$ MHz across the eight NV ensemble sensors), are decided experimentally by fitting the CW ODMR results in Fig. \ref{fig2}c. These results are found similar to those in \cite{foy2020wide}.

\subsubsection*{Machine learning methods for experimental magnetic localisation}

For magnetic localisation, we adopt a Convolutional Neural Network (CNN)-based machine-learning method for the magnetic inverse calculation. The machine-learning based method is suitable for scenarios where the magnetic field profile of the source is hard to characterise or often changing. For a different target (or change of the magnetic profile), one needs to retrain the model for optimal estimation precision. Also, certain physical bounds on the target movement have to apply (for example, movement only in the xy plane with fixed z), to avoid position redundancy that cannot be differentiated by magnetic fields. This needs to be more carefully considered especially when magnitude sensors are used (such as NV ensemble sensor based on nanodiamonds), compared to single-NV sensors, due to the lack of vectorial field information.

Here, we provide details on the CNN model used to achieve magnetic localisation in Fig. \ref{fig3} and Fig. \ref{fig4}. During data preprocessing, we fit the CW ODMR measurement results (after a two-point smoothing filter) by a double-gaussian-dip function, to extract the Zeeman splitting (difference of the two resonance frequencies) as well as the fitting standard deviation for the two resonances for all eight sensors. Note that we take the natural logarithm of these values so that the difference between sensors is more easily captured by the model. The labelled data is split such that 90\% is used for training, 7\% is used for validation while training, and 3\% is used for testing and visualisation in Fig. \ref{fig3}d. The CNN-based model is implemented using TensorFlow, starting with three Conv1D layers of 64, 128, and 256 dimensions, each followed by a ReLU activation function. Three Dense layers are then used (of 128, 64, and 2 dimensions) after flattening. An Adam optimiser is used with a learning rate of 0.005 and the model is trained for 3000 epochs under a batch size of 128. The training is monitored with a validation dataset and stopped before overfitting happens. This training can be easily done with a laptop in less than 30 minutes. Once trained, the same model can be used for magnetic localisation as long as experimental conditions are maintained (the magnetic needle tip setup and gesture). The magnetisation of the needle tip also remains throughout the experiments. See Supplementary Section B for details of the magnetic needle tip setup. 

More details on machine learning models used for the microrobot simulation can be found in the Supplementary Section D.

\backmatter


\section{Data availability}
The data that supports the plots within this paper and other findings of this study are available from the corresponding author upon reasonable request.

\section{Code availability}
The code used to generate the plots within this paper is available from the corresponding author upon reasonable request.


\bibliography{bibliography}

\section{Acknowledgments}
The authors acknowledge funding support from the Engineering and Physical Sciences Research Council (EPSRC) grant QC:SCALE EP/W006685/1. JAS acknowledges his EPSRC Quantum Technology Career Acceleration Fellowship (EP/C001220/1).

\end{document}